\pdfoutput=1

\documentclass[5p,times,twocolumn,number]{elsarticle}
\setcitestyle{open={[},close={]}}

\makeatletter
\def\ps@pprintTitle{%
  \let\@oddhead\@empty
  \let\@evenhead\@empty
  \let\@oddfoot\@empty
  \let\@evenfoot\@oddfoot}
\makeatother

\usepackage{babel}
\usepackage{graphbox}
\usepackage{multirow}
\usepackage{makecell}
\usepackage{amsmath,amssymb, gensymb}
\usepackage{hyperref}
\usepackage{url}
\usepackage{siunitx}
\usepackage{xspace}
\usepackage{microtype}
\usepackage{subfig}
\usepackage[useregional]{datetime2}
\usepackage{graphicx}
\usepackage{textcomp}
\usepackage{booktabs}
\usepackage{makecell} 
\usepackage{arydshln} 
\usepackage{comment}

\usepackage{pgfplots}
\pgfplotsset{compat=1.17}
\usetikzlibrary{patterns}

\begin{document}

\begin{frontmatter}
    
    \title{Electroluminescence Yield Measurements in Xenon Gas with the NEXT-DEMO++ Detector}

    \fntext[h]{Corresponding author}
\author[16]{J.~Renner\fnref{h}}
\author[16]{J.D.~Villamil}
\author[16]{N.~L\'opez-March}
\author[3]{K.~Mistry}
\author[16]{P.~Novella}
\author[16]{A.~Sim\'on}
\author[23]{V.~\'Alvarez}
\author[21]{J.M.~Benlloch-Rodr\'{i}guez}
\author[16,22]{M.~Cid}
\author[16]{C.~Cortes-Parra}
\author[23]{R.~Esteve}
\author[16]{F.~Kellerer}
\author[16]{J.~Mart\'in-Albo}
\author[23]{A.~Mart\'inez}
\author[16]{G.~Mart\'inez-Lema}
\author[16]{M.~Mart\'inez-Vara}
\author[16]{M.~Querol}
\author[16]{P.~Saharia}
\author[16]{M.~Sorel}
\author[16]{S.~Teruel-Pardo}

\author[21]{H.~Almaz\'an}
\author[7]{L.~Arazi}
\author[17]{I.J.~Arnquist}
\author[20]{F.~Auria-Luna}
\author[16]{S.~Ayet}
\author[22]{Y.~Ayyad}
\author[5]{C.D.R.~Azevedo}
\author[23]{F.~Ballester}
\author[21]{J.E.~Barcelon}
\author[21,9]{M.~del Barrio-Torregrosa}
\author[11]{F.I.G.M.~Borges}
\author[21,18]{A.~Brodoline}
\author[3]{N.~Byrnes}
\author[21]{A.~Castillo}
\author[17]{E.~Church}
\author[22]{X.~Cid}
\fntext[a]{Deceased.}
\author[11]{C.A.N.~Conde\fnref{a}}
\author[20]{F.P.~Coss\'io}
\author[15]{R.~Coupe}
\author[3]{E.~Dey}
\author[21]{P.~Dietz}
\author[21]{C.~Echeverria}
\author[21,9]{M.~Elorza}
\fntext[b]{Now at Weizmann Institute of Science, Israel.}
\author[7]{R.~Felkai\fnref{b}}
\author[12]{L.M.P.~Fernandes}
\fntext[c]{On leave.}
\author[21,8]{P.~Ferrario\fnref{c}}
\author[21]{P.~Ferrero Manche\~{n}o}
\author[4]{F.W.~Foss}
\author[19,8]{Z.~Freixa}
\author[23]{J.~Garc\'ia-Barrena}
\fntext[d]{NEXT Spokesperson.}
\author[21,8]{J.J.~G\'omez-Cadenas\fnref{d}}
\author[15]{J.W.R.~Grocott}
\author[15]{R.~Guenette}
\author[1]{J.~Hauptman}
\author[12]{C.A.O.~Henriques}
\author[22]{J.A.~Hernando~Morata}
\author[14]{P.~Herrero-G\'omez}
\author[23]{V.~Herrero}
\author[22]{C.~Herv\'es Carrete}
\author[7]{Y.~Ifergan}
\author[12]{A.F.B.~Isabel}
\author[3,15]{B.J.P.~Jones}
\author[21]{L.~Larizgoitia}
\author[20]{A.~Larumbe}
\author[6]{P.~Lebrun}
\author[21]{F.~Lopez}
\author[4]{R.~Madigan}
\author[12]{R.D.P.~Mano}
\author[20]{A.~Marauri}
\author[11]{A.P.~Marques}
\author[4]{R.L.~Miller}
\author[20]{J.~Molina-Canteras}
\author[21,8]{F.~Monrabal}
\author[12]{C.M.B.~Monteiro}
\author[23]{F.J.~Mora}
\author[3]{K.E.~Navarro}
\author[3]{D.R.~Nygren}
\author[21]{E.~Oblak}
\author[15]{I.~Osborne}
\author[10]{J.~Palacio}
\author[15]{B.~Palmeiro}
\author[6]{A.~Para}
\author[3]{I.~Parmaksiz}
\author[19]{A.~Pazos}
\author[21]{J.~Pelegrin}
\author[22]{M.~P\'erez Maneiro}
\author[20,21]{I.~Rivilla}
\author[18]{C.~Rogero}
\author[2]{L.~Rogers}
\fntext[e]{Now at University of North Carolina, USA.}
\author[21]{B.~Romeo\fnref{e}}
\fntext[f]{Now at Los Alamos National Laboratory, USA.}
\author[16]{C.~Romo-Luque\fnref{f}}
\author[10]{E.~Ruiz-Ch\'oliz}
\author[11]{F.P.~Santos}
\author[12]{J.M.F. dos~Santos}
\author[21,9]{M.~Seemann}
\author[14]{I.~Shomroni}
\author[5]{A.L.M.~Silva}
\author[12]{P.A.O.C.~Silva}
\author[21,8]{S.R.~Soleti}
\author[16]{J.~Soto-Oton}
\author[12]{J.M.R.~Teixeira}
\author[23]{J.F.~Toledo}
\author[21]{C.~Tonnel\'e}
\author[21]{S.~Torelli}
\author[21,13]{J.~Torrent}
\author[15]{A.~Trettin}
\author[21,19]{P.R.G.~Valle}
\author[4]{M.~Vanga}
\author[21,22]{P.~V\'azquez Cabaleiro}
\author[5]{J.F.C.A.~Veloso}
\author[15]{L.M.~Villar Padruno}
\author[15]{J.~Waiton}
\author[21,9]{A.~Yubero-Navarro}
\address[1]{
Department of Physics and Astronomy, Iowa State University, Ames, IA 50011-3160, USA}
\address[2]{
Argonne National Laboratory, Argonne, IL 60439, USA}
\address[3]{
Department of Physics, University of Texas at Arlington, Arlington, TX 76019, USA}
\address[4]{
Department of Chemistry and Biochemistry, University of Texas at Arlington, Arlington, TX 76019, USA}
\address[5]{
Institute of Nanostructures, Nanomodelling and Nanofabrication (i3N), Universidade de Aveiro, Campus de Santiago, Aveiro, 3810-193, Portugal}
\address[6]{
Fermi National Accelerator Laboratory, Batavia, IL 60510, USA}
\address[7]{
Unit of Nuclear Engineering, Faculty of Engineering Sciences, Ben-Gurion University of the Negev, P.O.B. 653, Beer-Sheva, 8410501, Israel}
\address[8]{
Ikerbasque (Basque Foundation for Science), Bilbao, E-48009, Spain}
\address[9]{
Department of Physics, Universidad del Pais Vasco (UPV/EHU), PO Box 644, Bilbao, E-48080, Spain}
\address[10]{
Laboratorio Subterr\'aneo de Canfranc, Paseo de los Ayerbe s/n, Canfranc Estaci\'on, E-22880, Spain}
\address[11]{
LIP, Department of Physics, University of Coimbra, Coimbra, 3004-516, Portugal}
\address[12]{
LIBPhys, Physics Department, University of Coimbra, Rua Larga, Coimbra, 3004-516, Portugal}
\address[13]{
Escola Polit\`ecnica Superior, Universitat de Girona, Av.~Montilivi, s/n, Girona, E-17071, Spain}
\address[14]{
Racah Institute of Physics, The Hebrew University of Jerusalem, Jerusalem 9190401, Israel}
\address[15]{
Department of Physics and Astronomy, Manchester University, Manchester. M13 9PL, United Kingdom}
\address[16]{
Instituto de F\'isica Corpuscular (IFIC), CSIC \& Universitat de Val\`encia, Calle Catedr\'atico Jos\'e Beltr\'an, 2, Paterna, E-46980, Spain}
\address[17]{
Pacific Northwest National Laboratory (PNNL), Richland, WA 99352, USA}
\address[18]{
Centro de F\'isica de Materiales (CFM), CSIC \& Universidad del Pais Vasco (UPV/EHU), Manuel de Lardizabal 5, San Sebasti\'an / Donostia, E-20018, Spain}
\address[19]{
Department of Applied Chemistry, Universidad del Pais Vasco (UPV/EHU), Manuel de Lardizabal 3, San Sebasti\'an / Donostia, E-20018, Spain}
\address[20]{
Department of Organic Chemistry I, Universidad del Pais Vasco (UPV/EHU), Centro de Innovaci\'on en Qu\'imica Avanzada (ORFEO-CINQA), San Sebasti\'an / Donostia, E-20018, Spain}
\address[21]{
Donostia International Physics Center, BERC Basque Excellence Research Centre, Manuel de Lardizabal 4, San Sebasti\'an / Donostia, E-20018, Spain}
\address[22]{
Instituto Gallego de F\'isica de Altas Energ\'ias, Univ.\ de Santiago de Compostela, Campus sur, R\'ua Xos\'e Mar\'ia Su\'arez N\'u\~nez, s/n, Santiago de Compostela, E-15782, Spain}
\address[23]{
Instituto de Instrumentaci\'on para Imagen Molecular (I3M), Centro Mixto CSIC - Universitat Polit\`ecnica de Val\`encia, Camino de Vera s/n, Valencia, E-46022, Spain}
%

    \begin{abstract}
The NEXT-DEMO++ detector, a high-pressure xenon gas time projection chamber serving as a prototype for the NEXT-100 experiment, was used to measure the electroluminescence (EL) yield as a function of reduced electric field ($E/p$) across pressures from 2.0 to 9.4 bar, utilizing the 41.5 keV de-excitation peak of $^{83m}$Kr. These measurements were made to examine the pressure dependence of the slope of the reduced EL yield $Y/p$, which has shown inconsistencies in the literature. The reduced yield was fitted with a linear model, revealing a modest ($\sim$5\%) change in slope, beginning around 5 bar and increasing with pressure up to 9.4 bar.
\end{abstract}

\end{frontmatter}

\section{Introduction}\label{s:intro}

High-pressure gaseous xenon time projection chambers (HPXe TPCs) represent a versatile and powerful technology in the field of particle detection, offering excellent energy resolution and topological reconstruction capabilities \cite{Bolotnikov:97, Ferrario:2015kta, NEXT:2020jmz, Nygren:2009zz}. Notable applications include the search for neutrinoless double beta decay ($\beta\beta0\nu$)~\cite{Avignone2008, Dolinski2019, Gomez-Cadenas:2023vca}, a process that, if observed, would imply lepton number violation and the Majorana nature of neutrinos~\cite{Bilenky_1987}.

An important characteristic of xenon-based gaseous detectors is the ability to employ electroluminescence (EL), also known as secondary scintillation, which provides proportional amplification of the ionization signal without the statistical fluctuations associated with avalanche multiplication \cite{Aprile:2008bga, freitas_2010, leardini_2022}. In EL-based TPCs, primary ionization electrons are drifted under a moderate electric field and then accelerated in a higher-field region to excite xenon atoms, producing vacuum ultraviolet (VUV) photons that can be detected by photosensors. This mechanism enables near-intrinsic energy resolution~\cite{Renner:2019pfe}, limited primarily by the low Fano factor~\cite{PhysRev.72.26} of xenon gas.

Previous studies of EL yield in high-pressure xenon have revealed dependencies on gas density and electric field strength. Measurements by Freitas et al. \cite{freitas_2010} indicated a non-linear increase in the EL yield with pressure in the range of 2--10 bar, suggesting potential deviations from the expected scaling. However, more recent work by Leardini et al. \cite{leardini_2022} did not observe such an enhancement, highlighting the need for further investigation. Understanding the pressure dependence of EL yield is essential for optimizing detector performance in high-pressure regimes.

The NEXT (Neutrino Experiment with a Xenon TPC) collaboration is developing HPXe TPC technology for the search for $\beta\beta0\nu$ in $^{136}$Xe. The NEXT-100 detector~\cite{NEXT:2025yqw}, a TPC capable of holding 100 kg of gas at 15 bar pressure, has achieved an energy resolution better than 1\% FWHM~\cite{NEXT_2025_energy} at the $Q_{\beta\beta}$ value of 2458 keV while providing topological discrimination to suppress backgrounds. As a prototype and technological demonstrator for NEXT-100, the NEXT-DEMO detector has been instrumental in validating key aspects of the design, including gas purity, electron drift parameters, and sensor performance \cite{Alvarez:2013gxa, Lorca:2014sra, Alvarez:2012xda}. In this work, we present measurements of the EL yield as a function of reduced electric field ($E/p$) in an upgraded NEXT-DEMO detector which we call NEXT-DEMO++. We use the de-excitation peak from $^{83m}$Kr as a reference and study xenon gas pressures from approximately 2 to 9 bar. These results aim to clarify the observed pressure-related variations and provide insights into the underlying scintillation mechanisms in dense xenon gas.

\section{The NEXT-DEMO++ Detector}\label{sec:detector}
\begin{figure}[!h]
    \centering
    \includegraphics[scale=0.42]{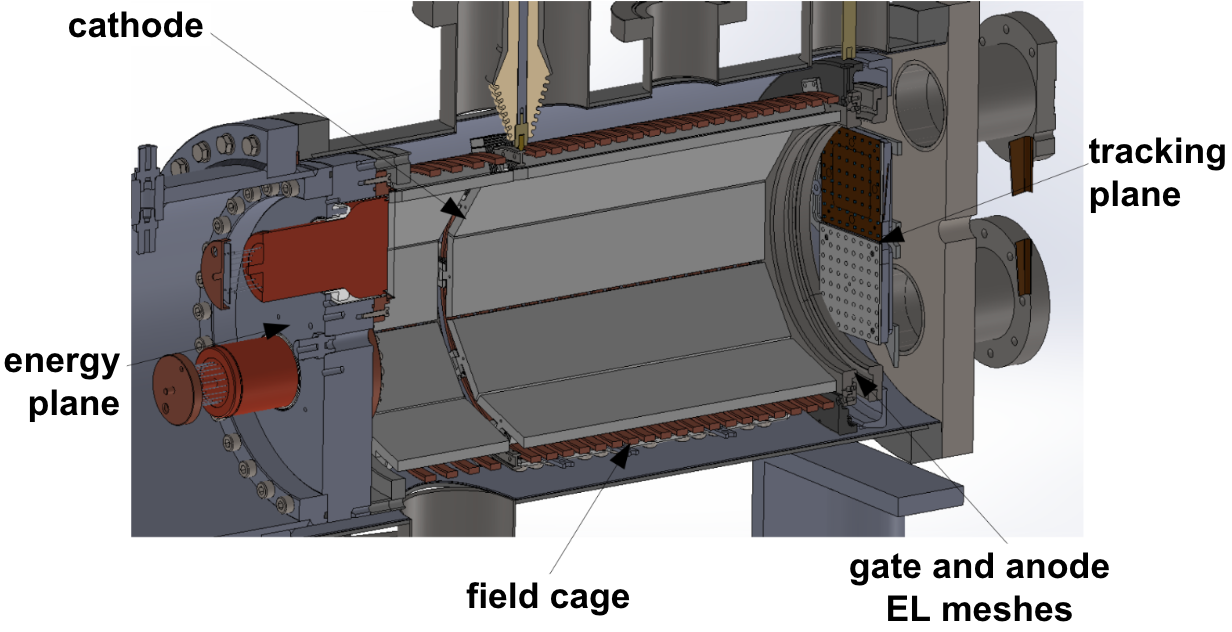}
    \caption{A drawing highlighting the key components of the NEXT-DEMO++ detector. The field cage consisting of copper rings enclosing reflective teflon panels surrounds the active region. The energy plane (3 PMTs) and tracking plane (256 SiPMs) observe the active region from opposite ends. Voltages placed on transparent conducting planes establish the electric fields: the cathode plane (wire grid not shown) is placed in front of the energy plane after a buffer region, and the gate and EL planes (meshes not shown) are located just in front of the tracking plane.}
    \label{fig:demo_CAD}
\end{figure}

NEXT-DEMO++ is housed in the same stainless-steel pressure vessel as NEXT-DEMO, capable of operating at pressures up to 15 bar, but several key internal components have been modified. A cross section of the upgraded detector is shown in Fig. \ref{fig:demo_CAD}.

The active volume of the TPC is a cylinder of approximately 30 cm in length and 20 cm in diameter. A reflective light tube, formed by ten polytetrafluoroethylene (PTFE) panels, encloses the active volume. The gas system ensures high purity through continuous recirculation via a hot getter (model SAES PS4-MT15-R-2), maintaining electron lifetimes exceeding 5 ms.

To create the electric fields necessary for detector operation, the volume is segmented by three wire grids: a cathode, a gate, and an anode. High voltage is supplied to the cathode and gate via custom feedthroughs. The three grids define three critical regions:

\emph{1. The buffer region:} a 10 cm buffer region that serves to degrade the high voltage set on the cathode to the grounded PMT-energy plane. The thickness of this region is set to establish a suitable reduced electric field, avoiding unwanted electroluminescence and mitigating neutral bremsstrahlung emission in this region~\cite{NEXT_PRX_2022}.

\emph{2. The drift region:} a 30 cm long volume between the cathode and the gate, where a moderate reduced electric field ($\sim$50 V/cm/bar) drifts ionization electrons. The field in this region is generated by the cathode-gate voltage difference, and is degraded along the field cage using metallic rings connected by resistors. The cathode consists of a grid of 15 parallel wires stretched over the circular cross section of the cylindrical field cage.

\emph{3. The electroluminescence region:} a 1 cm-long region of high reduced electric field (1-3 kV/cm/bar) between the gate and the anode, where light production occurs. The gate and anode are constructed from photoetched stainless steel meshes. These meshes were etched with hexagonal patterns of side length 2.5 mm and wire size 127 microns. They were tensioned to a force of approximately 2 kN. The gate is held at a negative voltage to establish the EL field, while the anode is grounded.

The primary scintillation emitted by xenon excitations produced during the creation of the ionization track are first observed, followed later in time by the EL photons. The detected primary scintillation is known as S1, while the detected EL photons are known as S2. Two distinct planes of photosensors are installed in the detector to observe the emitted photons. 

The first, known as the Energy Plane, is used to measure both S1 and S2. It is separated from the cathode by the buffer region, and is instrumented with 3 Hamamatsu R11410 photomultiplier tubes (PMTs), each with a diameter of 64 mm. As installed in NEXT-100~\cite{NEXT:2025yqw}, the PMTs are housed in individual copper enclosures, held at vacuum, and observe the full TPC volume through sapphire windows coated with a layer of poly(3,4-ethylenedioxythiophene) (PEDOT), a transparent compound which assists in grounding the energy plane and preventing charge-up.

The second plane of photosensors, the Tracking Plane, in practice measures only S2. It is positioned directly behind the anode mesh and consists of an array of 256 Hamamatsu S13372-1350TE silicon photomultipliers (SiPMs) arranged on 4 kapton boards containing 64 sensors each. The sensors cover an area of 150 x 150 mm$^2$ in X-Y. A PTFE mask is placed over each kapton board to improve the reflectivity of the plane. 

The PTFE panels of the light tube surrounding the active region, the sapphire windows of the PMT enclosures and the SiPMs and their masks are all coated with tetraphenyl butadiene (TPB). This compound shifts the xenon vacuum ultraviolet (VUV) scintillation light ($\lambda\sim 172$ nm) to the blue spectrum ($\lambda\sim 430$ nm) \cite{NEXT_TPB_2012}. As the SiPMs are not sensitive to VUV light, and the PMT detection efficiency is much lower in the VUV range, this significantly improves overall light collection efficiency.

\section{Data summary}\label{sec:dataruns}

The data presented in this work were acquired during a dedicated experimental campaign from April to June of 2025 using an internal $^{83m}$Kr source. The $^{83m}$Kr isotope, produced from the decay of $^{83}$Rb embedded in a resin bead placed within the detector's gas system, provides monoenergetic de-excitations at approximately 41.5 keV, ideal for measuring the EL yield due to the resulting point-like energy depositions uniformly distributed throughout the active volume.

\begin{figure*}[!ht]
    \centering
    \includegraphics[scale=0.6]{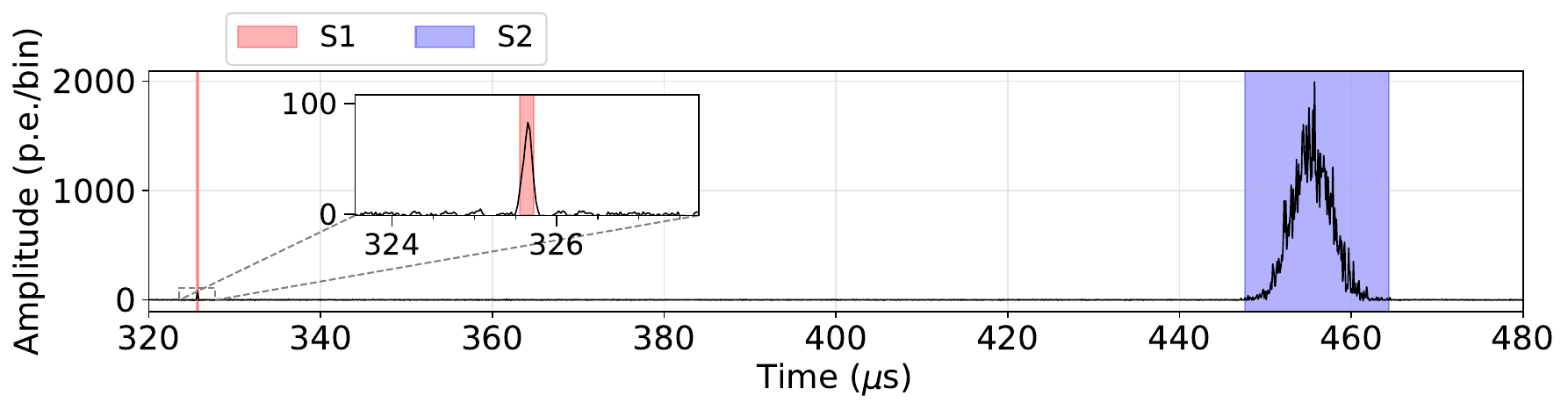}
    \caption{An example summed PMT waveform for a candidate $^{83m}$Kr event, showing the regions corresponding to S1 (primary scintillation) and S2 (electroluminescence).}
    \label{fig:waveform}
\end{figure*}

We report data taken starting at 2.0~bar and increasing in approximately 1~bar increments up to 9.4~bar. Note that though the pressure vessel is rated for operation up to 15 bar, the gas compressor employed in this study only allowed for operation up to a maximum of 10~bar. Pressure was measured using a WIKA WUD-20 pressure sensor with 0--25 bar range. At each pressure point $p$, a series of measurements was taken by systematically varying the EL field, $E$, to scan a range of $E/p$ values. The cathode voltage was adjusted to maintain a constant reduced drift field of approximately 50~V\,cm$^{-1}$\,bar$^{-1}$, yielding a drift velocity of 0.87 mm/$\mu$s, though the final two points of the 9.4~bar data were taken at slightly lower reduced drift fields (45 and 39~V\,cm$^{-1}$\,bar$^{-1}$) due to limitations of the cathode high voltage power supply. Operating temperatures increased with increasing pressure from 17.1\degree C (290.3 K) at 2 bar to 21.3\degree C (294.4 K) at 9 bar, a 1.4\% increase. The operational parameters for each pressure are summarized in Table~\ref{tab:run_summary}.

\begin{table}[h!]
    \centering
    \caption{Summary of the operational conditions for the EL yield scan campaign. For each pressure studied, the table lists the number of distinct $E/p$ points analyzed and the range of the $E/p$ values scanned.}
    \label{tab:run_summary}
    \begin{tabular}{cccc}
    \hline
    \hline
    \textbf{Pressure (bar)} & \textbf{Data Points} & \textbf{E/p Range (kV/cm/bar)} \\
    \hline
    2.0052 $\pm$ 0.0036 & 9  & 1.2 -- 2.8 \\
    3.0690 $\pm$ 0.0036 & 10 & 1.2 -- 3.0 \\
    4.0622 $\pm$ 0.0034 & 10 & 1.2 -- 3.0 \\
    5.1604 $\pm$ 0.0034 & 10 & 1.2 -- 3.0 \\
    6.1571 $\pm$ 0.0036 & 10 & 1.2 -- 3.0 \\
    7.2177 $\pm$ 0.0057 & 9  & 1.2 -- 3.0 \\
    8.2689 $\pm$ 0.0040 & 9  & 1.2 -- 2.8 \\
    9.4157 $\pm$ 0.0040 & 8  & 1.2 -- 2.6 \\
    \hline
    \hline
    \end{tabular}
\end{table}

The sensor planes were calibrated via a multi-Gaussian fit to the noise and single-photoelectron (SPE) peaks of each sensor in dedicated calibration runs. In the case of the PMTs, an LED installed inside the detector on the tracking plane was pulsed every 50 microseconds to produce SPE events, and the resulting spectrum was fit to a single function consisting of several Gaussians. Dark counts were used to calibrate the SiPMs, and the noise and SPE peak were fit separately with two different Gaussian functions. The SPE gain values obtained for the PMTs were found to be 34.6, 34.0, and 41.5 ADC units/p.e, and for the SiPMs varied between 16-19 ADC units/p.e. These values were not found to vary appreciably during the data-taking campaign.

To promote stable detector conditions for each pressure setting, all runs for a given pressure were performed consecutively, on the same day or on 2 consecutive days. Each individual run consisted of acquiring $\sim 10^5$ triggered events. The trigger was set with minimum and maximum charge thresholds on each of the 3 PMTs to select integrated S2 values consistent with the $^{83m}$Kr peak. A 900 $\mu$s buffer was saved for each event, including 450 $\mu$s pre-trigger and 450 $\mu$s post-trigger, so that more than an entire drift length (approx. 350 $\mu$s for a 31 cm drift length at electron drift velocity of 0.87 mm/$\mu$s) of information was available before and after the trigger arrival time. An example waveform of a candidate Kr event is shown in Figure \ref{fig:waveform}. A longer run of $> 10^6$ events was taken for each pressure to characterize stability over a longer time period. Maximum variations in light yield of order 1.5\% were observed, which have been accounted for as systematic errors in the measured light yields.

The electron lifetime, indicative of gas purity and attachment losses, was monitored via the dependence of the S2 signal on drift time. Values ranged from approximately 7--30 ms throughout the campaign, while for a given pressure the lifetime was found to remain constant to within $\sim$25\%. Considering the 350 $\mu$s drift time, a (conservative) $\sim$25\% variation on a mean lifetime of 20 ms would correspond to at most a variation of $\exp(-350/25000) - \exp(-350/15000) \approx 0.9$\% in light yield measurements. As this is a conservative estimate and is likely correlated with the $\sim$1.5\% variations in light yield observed over time, we maintain an overall systematic error on the measured yields of 1.5\%.

\section{Analysis}\label{s:analysis}

\subsection{Event selection and reconstruction}
Data processing was performed using the Invisible Cities (IC) software framework \cite{IC}, a modular Python-based package developed by the NEXT collaboration for the analysis of high-pressure xenon TPC data. IC handles the full reconstruction chain from raw waveforms to high-level event parameters.

The processing began with analysis of the PMT and SiPM waveforms, accounting for pedestal subtraction and sensor-specific gains. Peak finding was then applied to the summed, calibrated PMT waveform to identify S1 and S2 peaks. Integration of the S2 peak gave a measurement of the energy of the event. The average S2 time, computed by weighting the time since S1 of each S2 time bin in the waveform by its charge, gave an average drift time of the event, corresponding to a drift distance Z. Individual SiPM waveforms were integrated within the S2 time window, and an average position was computed by weighting the (X,Y) of each SiPM by its integrated charge. 

\begin{figure}[!h]
    \centering
    \includegraphics[scale=0.4]{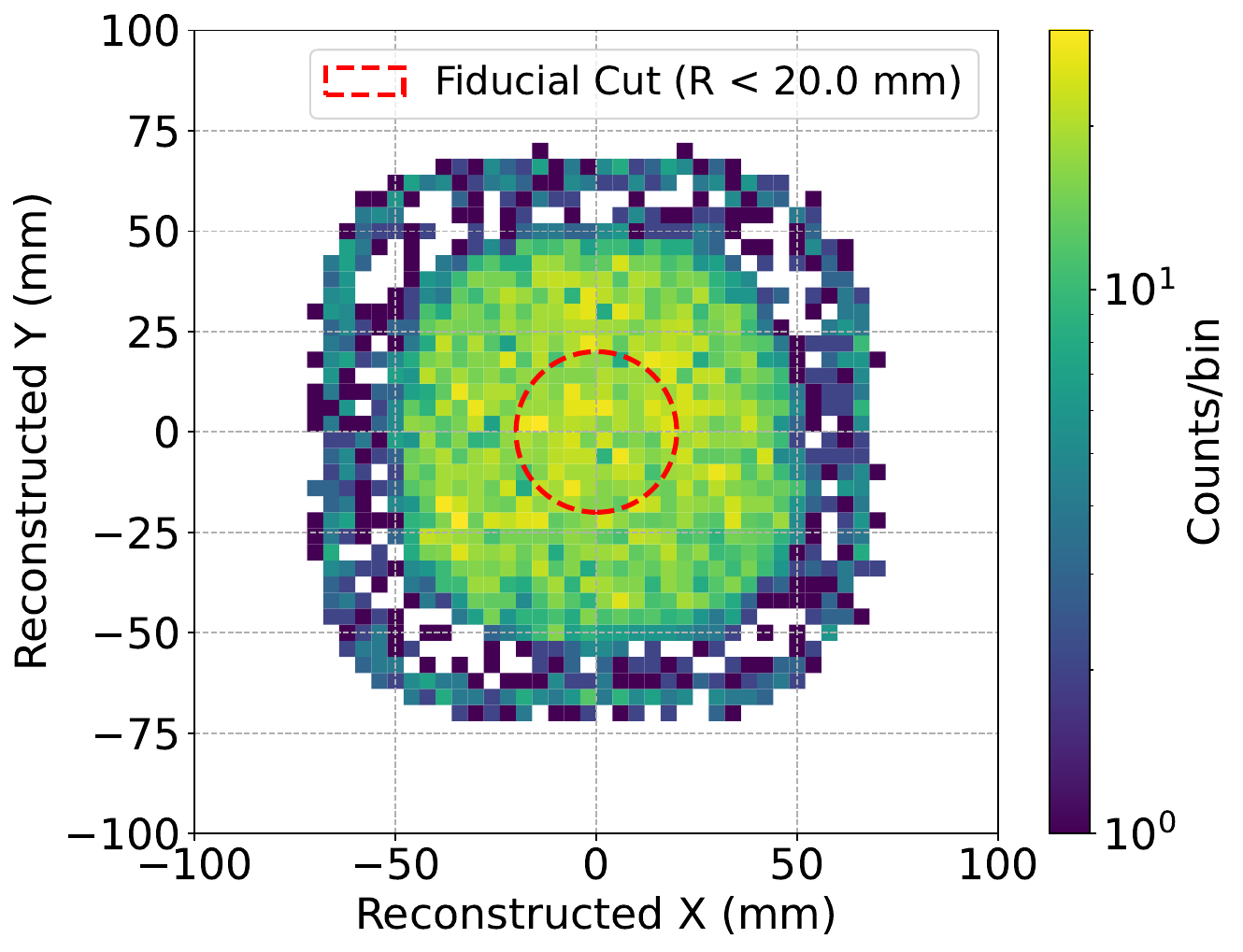}
    \includegraphics[scale=0.36]{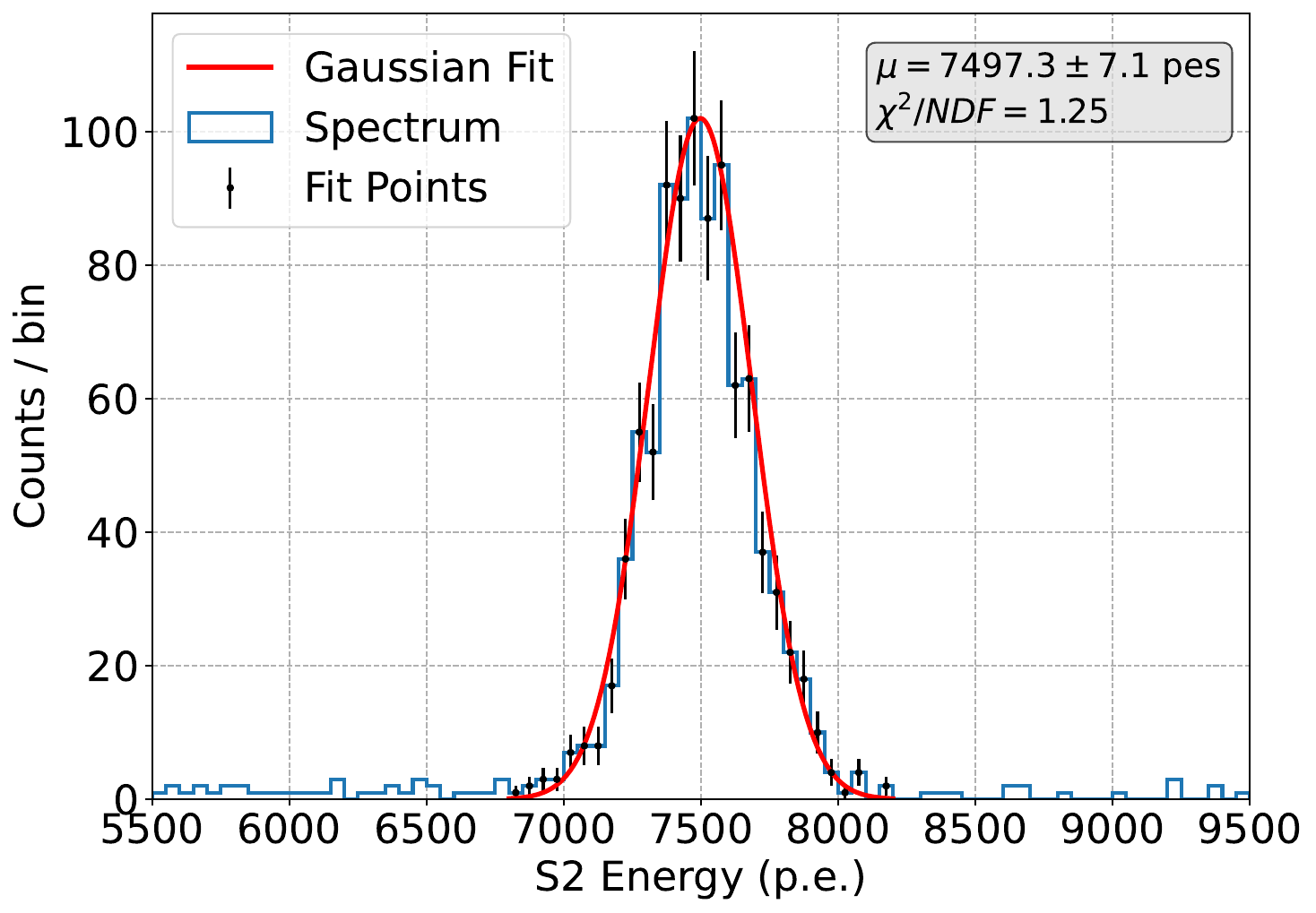}
    \caption{$^{83m}$Kr peak analysis for a single run at 5 bar. (Top) XY distribution of reconstructed $^{83m}$Kr events and the applied fiducial cut $R < 20$ mm. The distribution reflects the uniform illumination by the Kr source, with reductions near the edges due to partial SiPM coverage of the tracking plane (-75 mm $<$ X,Y $<$ 75 mm) and due to reconstruction effects. (Bottom) Gaussian fit to the S2 energy spectrum for fiducial $^{83m}$Kr events. The mean value corresponds to the EL-amplified signal for 41.5 keV.}
    \label{fig:kr_analysis}
\end{figure}

Event selection was focused on isolating $^{83m}$Kr events. Criteria included exactly one S1 peak with energy consistent with the Kr primary scintillation, exactly one S2 peak, drift time between 100 and 300 $\mu$s, and a tight fiducial cut, $R < 20$ mm, in the transverse plane to minimize edge effects and non-uniform regions. These fiducial cuts in (X,Y) and Z (drift time) served to minimize energy variations due to location within the detector, as no corrections for lifetime or geometric effects were applied. Figure \ref{fig:kr_analysis} (top) shows the X-Y distribution of reconstructed events for a representative run (at 5 bar), along with the applied fiducial cut.

Following the above selection, the S2 energy spectrum was constructed from the summed PMT charge in the S2 peak (in photoelectrons). The spectrum was fit with a Gaussian function to extract the mean S2 yield for the 41.5 keV Kr peak. Figure \ref{fig:kr_analysis} (bottom) shows an example fit for the same representative run. 

\subsection{Electroluminescent yield}
To translate the experimentally determined $^{83m}$Kr yields into an EL yield per electron, we computed the number of electron-ion pairs $N_e$ produced by an incident electron depositing energy $E_{\text{dep}}$ within the active volume,

\begin{equation}
    N_e = \frac{E_{\text{dep}}}{W_{\text{i}}},
    \label{eq:initial_electrons}
\end{equation}

\noindent where $W_{\text{i}}$ is the average energy required to create an electron-ion pair in xenon gas. The applied reduced drift field (50 V/cm/bar) was set high enough to reduce fluctuations due to electron-ion recombination to a negligible level~\cite{Bolotnikov_1993}. The yield $Y$, in photons/e-, of the EL amplification of the ionization charge is well-described as a function of the reduced electric field, $E/p$, in kV/cm/bar, and EL gap distance $d$ in cm~\cite{Aprile:2008bga,Nygren:2009zz},

\begin{equation}
    Y = A\left(\frac{E}{p} - X_0\right) p d.
    \label{eq:total_yield}
\end{equation}

\noindent Here $A$ is the intrinsic EL gain coefficient of the gas in photons/e$^{-}$/kV and $X_0$ is the reduced electric field threshold in kV/cm/bar. Generally accepted empirical values of these parameters have been determined to be~\cite{Aprile:2008bga} $A = 140$ photons/e$^{-}$/kV and $X_0 = 0.83$ kV/cm/bar. Note that this formalism expresses the EL yield as a function of pressure $p$ while the actual dependence is on gas density. As our temperature variations were observed to be low (of order 1\% over the entire data taking period) and did not significantly affect the proportionality between pressure and density, we consider this to be a valid approximation.

We can compute $Y$ from our data given our fit mean $^{83m}$Kr peak S2 values, $\overline{S_2}$ as 

\begin{equation}
Y = \frac{\overline{S_\text{2}}}{\epsilon_{LC} \cdot N_e},
\label{eqn:yield}
\end{equation}

\noindent where $\epsilon_{LC}$ is the light collection efficiency (accounting for geometric coverage and quantum efficiency), and $N_{\text{e}} = E_{\rm Kr} / W_i$ is the number of primary electrons produced by the Kr de-excitation with $E_{\rm Kr} = 41.56$ keV and ionization work function $W_i = 21.9 \pm 0.4$ eV~\cite{Platzman1961}. 

Due to the complexity of modeling all geometric effects that influence  light collection in the detector, and due to the large uncertainty in reported values of $W_{sc}$, the average energy required to create a primary scintillation photon in xenon gas, it is difficult to determine $\epsilon_{LC}$ precisely. It was computed by first performing a linear fit to the 2D distribution of the number of primary scintillation photons detected for a $^{83m}$Kr event vs. drift distance $Z$, and taking the intercept of the fit, $S_{1}^{0}$, at $Z=0$. The intercept was found to be $S_{1}^{0} = 4.33 \pm 1.01$. As the S1 vs. Z band was quite wide, the error on this value is dominated by systematics of the linear fit. The expected number of photons produced was computed as $N_{ph} = E_{\text{Kr}}/W_{sc}$. The light collection efficiency was then computed as $\epsilon_{LC} = S_{1}^{0}/N_{ph}$. Reported values of $W_{sc}$ vary considerably, from 34-111 eV depending on the evaluated particle type (alpha, x-ray, or gamma), and we chose $W_{sc} = 39.9 \pm 8$ \cite{Henriques_2024}, resulting in a final value of $\epsilon_{LC} = (0.45 \pm 0.14)$\%.

\begin{figure}[!htb]
    \centering
    \includegraphics[scale=0.39]{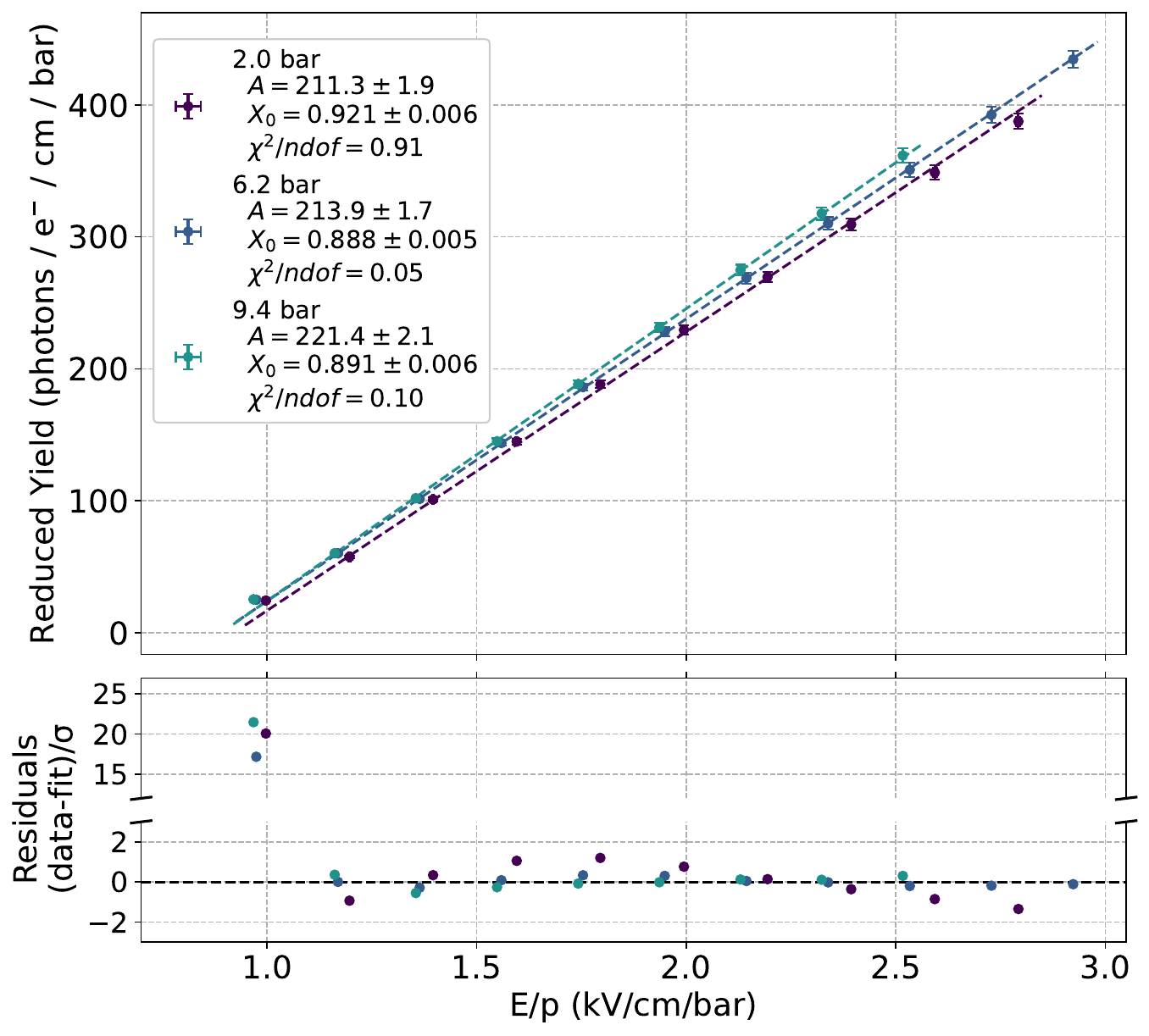}
    \caption{Reduced electroluminescence yield $Y/(p\cdot d)$ as a function of reduced electric field $E/p$ for three different pressures. For each pressure, the reduced yield values are fit to a line to extract the corresponding values of $A$ and $X_0$ (Eqn. \ref{eq:total_yield}). Note that only E/p values greater than 1.1 kV/cm/bar were included in the fit, though residuals are shown for all data points, highlighting the discrepancy observed at the lowest E/p values. Note also that the absolute values obtained for the fit parameter $A$ are significantly different than the accepted theoretical value ($\sim$140 photons/e$^{-}$/kV), and this is most likely due to the large uncertainty in the computed light collection efficiency.}
    \label{fig:scan_summary}
\end{figure}

\begin{figure*}[!htb]
    \centering
    \includegraphics[scale=0.6]{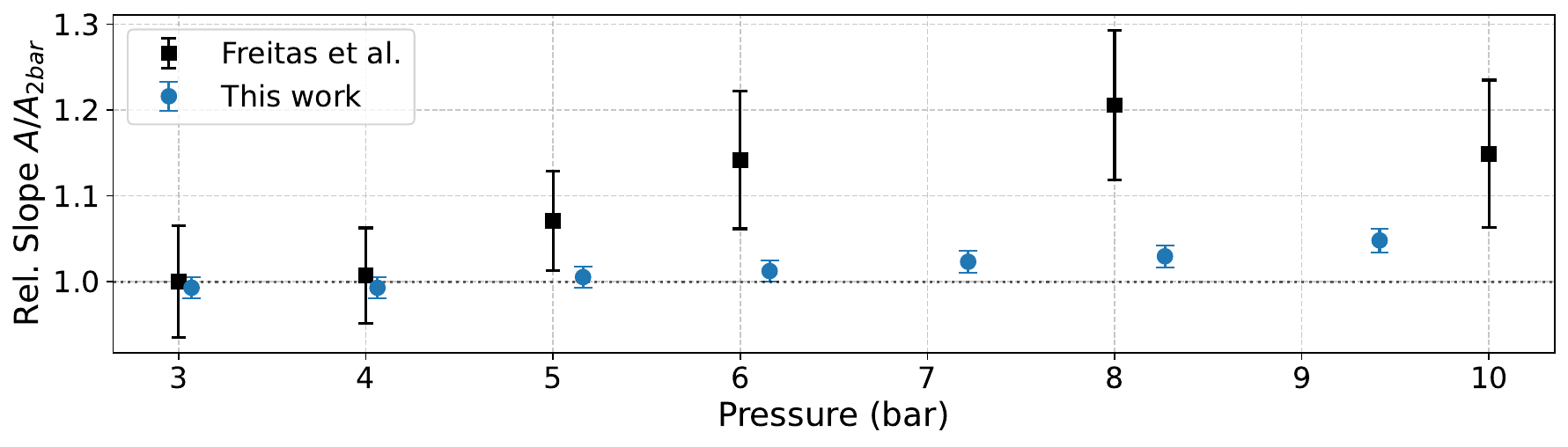}
    \includegraphics[scale=0.6]{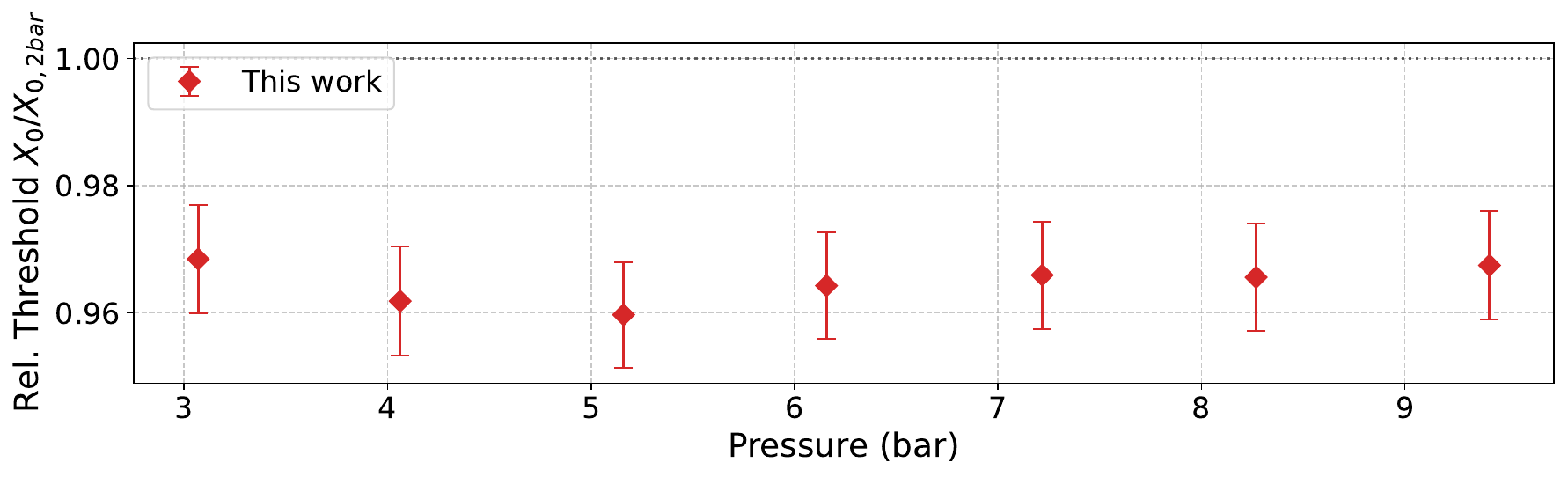}
    \caption{Fit parameters $A$ (above) and $X_0$ (below) for each yield vs. E/p curve (see Fig. \ref{fig:scan_summary}) for all pressures studied, shown relative to the values obtained at 2 bar. The relative slope values from Freitas et al. were computed using the measurements listed in Table 1 of the corresponding reference~\cite{freitas_2010}.}
    \label{fig:fit_parameters}
\end{figure*}

Once the yield $Y$ was computed for each pressure, the reduced differential yield $Y/(p\cdot d)$ could then be fitted as a function of the reduced electric field $E/p$. The functional form from Eqn. \ref{eq:total_yield} is $Y/({p\cdot d}) = A \left( E/p - X_0 \right)$. Uncertainties include statistical errors from the S2 peak fits (which are relatively small, see Fig. \ref{fig:kr_analysis}) and the 1.5\% systematic error from temporal variations, added in quadrature. Figure \ref{fig:scan_summary} shows the fitted differential yield versus $E/p$ curves for three representative pressures. The raw data for all pressures used to compute these curves is recorded in Table \ref{tab:el_scan_summary_dense}. Values for $E/p \approx 1.0$ kV/cm/bar are recorded, but as they significantly deviated from the principal linear trend for each pressure, likely due to effects near the EL threshold that our linear model does not correctly explain, they were not included in the fit. The fit parameters for all pressure scans are summarized in Fig. \ref{fig:fit_parameters}, showing the slope $A$ and threshold $X_0$ as functions of pressure. The obtained $\chi^{2}/N_{\mathrm{dof}}$ values from the fits varied from 0.03 to 0.91, indicating that our assigned systematic error was conservative in most cases.

We note that though the relative errors on the yield values and fit parameters in figures \ref{fig:scan_summary} and \ref{fig:fit_parameters} are on the order of several percent, the absolute values of these parameters all incur a $\sim$30\% overall systematic error due to the large uncertainty in the light collection efficiency $\epsilon_{LC}$. Errors on other key quantities, such as the EL gap width (estimated to be $\sim$0.15 mm) and the absolute pressure value within a single scan, are small in comparison.

By considering the EL yield as a function of $E/p$, we have assumed in this analysis that xenon behaves as an ideal gas. While it is common to refer to the reduced electric field in units of $E/p$, the physics behind the EL yield actually depends on $E/N$, where $N$ is the xenon gas density. As pressure $p$ (or density $N$) cancels out in the first term of Eqn. \ref{eqn:yield}, this approximation affects only the threshold $X_0$. When accounting for non-ideal effects in the xenon pressure-density relation, the decrease in $X_0$ is found to be approximately 7\% (from 0.92 to 0.85 kV/cm/bar), still within the experimental errors.
\section{Discussion and conclusions}\label{s:discussion}
The present measurements of EL yield reveal a mild increase in the slope $A$ with pressure. The trend begins at pressures greater than 5 bar, rising from approximately $A=211$ to $A=221$ photons/e$^{-}$/kV at 9 bar, corresponding to a $\sim$5\% increase over the range. No substantial increase is observed until pressures greater than 5 bar, and the significance of a non-zero effect, assuming a linear model for the increase in relative slope from Figure \ref{fig:fit_parameters} (top), is found to be 3.7$\sigma$. We also observe a slight decrease in the EL threshold $X_0$, most notably between 2-3 bar. We have confirmed that the increase in slope is still present even if the linear fits to the EL curves are done with a fixed threshold, ruling out the possibility that the increase in slope is solely due to correlation with the threshold parameter in the fit procedure. Our findings of an upward trend in slope at higher pressures follow the same direction as the effect reported by Freitas et al. \cite{freitas_2010}, although the magnitude of the effect observed in this study ($\sim 5$\%) is less than observed ($\sim 20$\%) in that work. Leardini et al. \cite{leardini_2022} reported no significant pressure-dependent variation in yield across 3--10 bar. The raw data on which our analysis is based, including EL voltages and mean values of the fits to the $^{83m}$Kr peak in photoelectrons are tabulated in Table \ref{tab:el_scan_summary_dense}. 

The small increase in slope $A$ as well as the observed decrease in the EL threshold $X_0$ may be due to detector-related effects. We explored the possibility of mechanical bending of the EL meshes under high voltage, which could enhance the yield at higher pressures. However, an analytical calculation~\cite{NEXT:2023blw} of the mechanical bending of the DEMO mesh shows very little mesh displacement ($< 0.4$ mm), even with an electric field near the maximum EL field applied in this study (25 kV/cm) and mesh tension well below the possible values in our setup (50 N), indicating that significant deflection is unlikely at the applied voltages.

We also evaluated the possibility of enhanced EL production in the high electric fields near the mesh wires. However, a simulation of EL yield with varying $E/p$ in a COMSOL-based field model of the employed meshes did not show a significant increase in the slope of EL yield vs. $E/p$ with increasing pressure. 

Additional light production mechanisms may also contribute to the observed variations in EL scintillation parameters. Non-linear VUV ionization of the mesh material yielding additional electrons traversing the EL gap could result in an apparent increase in reduced EL yield given the higher electric fields and absolute yields at higher pressures. Other delayed light production mechanisms such as VUV-induced fluorescence of quartz present in the windows of photosensors~\cite{Sorensen_2025} could produce light over longer time scales. Despite these possibilities, we do not observe consistent signs of additional light production in the acquired waveforms.

\begin{table*}[h!tb]
\centering
\caption{Data for EL scans at all pressures studied. Each cell contains the EL gate voltage (V), the fitted Kr peak mean $\overline{S_\text{2}}$ (pes), and its statistical uncertainty (pes). To compute the exact E/p in kV/cm/bar in each case, convert the gate voltage values to kV and divide by the pressure and EL gap width $d = 1.0$ cm. To compute the reduced differential yield and errors (statistical), use Eqn. \ref{eqn:yield} with $\epsilon_{{LC}} = 0.0045$, and $N_{e} = E_{{\rm Kr}} / W_i$ with $E_{{\rm Kr}} = 41.56$ keV and $W_i = 21.9$ eV, and divide by the pressure (in bar) and EL gap width $d = 1.0$ cm.}
\label{tab:el_scan_summary_dense}
\setlength{\tabcolsep}{5pt} 
\renewcommand{\arraystretch}{1.5} 
\begin{tabular}{l|c:c:c:c:c:c:c:c:c:c:c}
\hline\hline
 & \multicolumn{11}{c}{\textbf{Nominal E/p (kV/cm/bar)}} \\
\textbf{{p (bar)}} & 1.0 & 1.2 & 1.4 & 1.6 & 1.8 & 2.0 & 2.2 & 2.4 & 2.6 & 2.8 & 3.0 \\
\hline
2.0052 & \makecell{2000V \\ 386 \\ $\pm$3} & \makecell{2400V \\ 909 \\ $\pm$2} & \makecell{2800V \\ 1597 \\ $\pm$1} & \makecell{3200V \\ 2292 \\ $\pm$2} & \makecell{3600V \\ 2976 \\ $\pm$3} & \makecell{4000V \\ 3631 \\ $\pm$4} & \makecell{4400V \\ 4265 \\ $\pm$4} & \makecell{4800V \\ 4896 \\ $\pm$6} & \makecell{5200V \\ 5518 \\ $\pm$4} & \makecell{5600V \\ 6131 \\ $\pm$3} & -- \\ \hdashline
3.0690 & \makecell{3000V \\ 560 \\ $\pm$4} & \makecell{3600V \\ 1411 \\ $\pm$4} & \makecell{4200V \\ 2425 \\ $\pm$5} & \makecell{4800V \\ 3456 \\ $\pm$5} & \makecell{5400V \\ 4478 \\ $\pm$4} & \makecell{6000V \\ 5476 \\ $\pm$5} & \makecell{6600V \\ 6433 \\ $\pm$6} & \makecell{7200V \\ 7401 \\ $\pm$5} & \makecell{7800V \\ 8331 \\ $\pm$7} & \makecell{8400V \\ 9237 \\ $\pm$5} & \makecell{9000V \\ 10157 \\ $\pm$10} \\ \hdashline
4.0622 & \makecell{4000V \\ 810 \\ $\pm$2} & \makecell{4800V \\ 1969 \\ $\pm$4} & \makecell{5600V \\ 3318 \\ $\pm$4} & \makecell{6400V \\ 4692 \\ $\pm$5} & \makecell{7200V \\ 6033 \\ $\pm$5} & \makecell{8000V \\ 7358 \\ $\pm$4} & \makecell{8800V \\ 8650 \\ $\pm$5} & \makecell{9600V \\ 9931 \\ $\pm$7} & \makecell{10400V \\ 11189 \\ $\pm$8} & \makecell{11200V \\ 12444 \\ $\pm$5} & \makecell{12000V \\ 13706 \\ $\pm$7} \\ \hdashline
5.1604 & \makecell{5000V \\ 981 \\ $\pm$3} & \makecell{6000V \\ 2398 \\ $\pm$4} & \makecell{7000V \\ 4087 \\ $\pm$4} & \makecell{8000V \\ 5806 \\ $\pm$5} & \makecell{9000V \\ 7497 \\ $\pm$7} & \makecell{10000V \\ 9167 \\ $\pm$8} & \makecell{11000V \\ 10807 \\ $\pm$8} & \makecell{12000V \\ 12453 \\ $\pm$9} & \makecell{13000V \\ 14069 \\ $\pm$8} & \makecell{14000V \\ 15717 \\ $\pm$9} & \makecell{15000V \\ 17383 \\ $\pm$12} \\ \hdashline
6.1571 & \makecell{6000V \\ 1213 \\ $\pm$3} & \makecell{7200V \\ 2922 \\ $\pm$4} & \makecell{8400V \\ 4925 \\ $\pm$5} & \makecell{9600V \\ 6980 \\ $\pm$6} & \makecell{10800V \\ 9041 \\ $\pm$7} & \makecell{12000V \\ 11072 \\ $\pm$8} & \makecell{13200V \\ 13055 \\ $\pm$11} & \makecell{14400V \\ 15066 \\ $\pm$7} & \makecell{15600V \\ 17044 \\ $\pm$11} & \makecell{16800V \\ 19067 \\ $\pm$14} & \makecell{18000V \\ 21110 \\ $\pm$14} \\ \hdashline
7.2177 & \makecell{7000V \\ 1410 \\ $\pm$4} & \makecell{8400V \\ 3383 \\ $\pm$5} & \makecell{9800V \\ 5733 \\ $\pm$5} & \makecell{11200V \\ 8149 \\ $\pm$9} & \makecell{12600V \\ 10532 \\ $\pm$7} & \makecell{14000V \\ 12944 \\ $\pm$10} & \makecell{15400V \\ 15300 \\ $\pm$10} & \makecell{16800V \\ 17722 \\ $\pm$11} & -- & \makecell{19600V \\ 22455 \\ $\pm$15} & \makecell{21000V \\ 24908 \\ $\pm$33} \\ \hdashline
8.2689 & \makecell{8000V \\ 1600 \\ $\pm$4} & \makecell{9600V \\ 3876 \\ $\pm$5} & \makecell{11200V \\ 6532 \\ $\pm$6} & \makecell{12800V \\ 9330 \\ $\pm$8} & \makecell{14400V \\ 12096 \\ $\pm$7} & \makecell{16000V \\ 14843 \\ $\pm$12} & \makecell{17600V \\ 17597 \\ $\pm$10} & \makecell{19200V \\ 20321 \\ $\pm$14} & \makecell{20800V \\ 23106 \\ $\pm$19} & \makecell{22400V \\ 25909 \\ $\pm$22} & -- \\ \hdashline
9.4157 & \makecell{9115V \\ 1874 \\ $\pm$3} & \makecell{10938V \\ 4475 \\ $\pm$5} & \makecell{12761V \\ 7572 \\ $\pm$6} & \makecell{14584V \\ 10777 \\ $\pm$8} & \makecell{16407V \\ 13987 \\ $\pm$10} & \makecell{18230V \\ 17185 \\ $\pm$12} & \makecell{20053V \\ 20412 \\ $\pm$15} & \makecell{21876V \\ 23597 \\ $\pm$16} & \makecell{23699V \\ 26865 \\ $\pm$27} & -- & -- \\
\hline\hline
\end{tabular}
\end{table*}

\section*{Acknowledgments}
The NEXT Collaboration acknowledges support from the following agencies and institutions: the European Research Council (ERC) under Grant Agreemenst No. 951281-BOLD and 101039048-GanESS; the European Union’s Framework Programme for Research and Innovation Horizon 2020 (2014–2020) under Grant Agreement No. 860881-HIDDeN; the MCIN/AEI of Spain and ERDF A way of making Europe under grants PID2021-125475NB and RTI2018-095979, and the Severo Ochoa and Mar\'ia de Maeztu Program grants CEX2023-001292-S, CEX2023-001318-M and CEX2018-000867-S; the Generalitat Valenciana of Spain under grants PROMETEO/2021/087, ASFAE/2022/028, ASFAE/2022/029, CISEJI/2023/27 and CIDEXG/2023/16; the Department of Education of the Basque Government of Spain under the predoctoral training program non-doctoral research personnel; the Spanish la Caixa Foundation (ID 100010434) under fellowship code LCF/BQ/PI22/11910019; the Portuguese FCT under project UID/FIS/04559/2020 to fund the activities of LIBPhys-UC; the Israel Science Foundation (ISF) under grant 1223/21; the Pazy Foundation (Israel) under grants 310/22, 315/19 and 465; the US Department of Energy under contracts number DE-AC02-06CH11357 (Argonne National Laboratory), DE-AC02-07CH11359 (Fermi National Accelerator Laboratory), DE-FG02-13ER42020 (Texas A\&M), DE-SC0019054 (Texas Arlington) and DE-SC0019223 (Texas Arlington); the US National Science Foundation under award number NSF CHE 2004111; the Robert A Welch Foundation under award number Y-2031-20200401. Finally, we are grateful to the Laboratorio Subterr\'aneo de Canfranc for hosting and supporting the NEXT experiment. The authors used the Google Gemini AI tool to assist in the elaboration and construction of the manuscript; all generated content was reviewed and edited by the authors, who assume full responsibility for the final version.

\bibliographystyle{elsarticle-num}
\bibliography{NextRefs}

\begin{thebibliography}{10}
\expandafter\ifx\csname url\endcsname\relax
  \def\url#1{\texttt{#1}}\fi
\expandafter\ifx\csname urlprefix\endcsname\relax\def\urlprefix{URL }\fi
\expandafter\ifx\csname href\endcsname\relax
  \def\href#1#2{#2} \def\path#1{#1}\fi

\bibitem{Bolotnikov:97}
A.~Bolotnikov, B.~Ramsey, The spectroscopic properties of high-pressure xenon,
  Nucl. Inst. Meth. A 396 (1997) 360--370.
\newblock \href {https://doi.org/10.1016/S0168-9002(97)00784-5}
  {\path{doi:10.1016/S0168-9002(97)00784-5}}.

\bibitem{Ferrario:2015kta}
P.~Ferrario, et~al., {First proof of topological signature in the high pressure
  xenon gas TPC with electroluminescence amplification for the NEXT
  experiment}, JHEP 01 (2016) 104.
\newblock \href {http://arxiv.org/abs/1507.05902} {\path{arXiv:1507.05902}},
  \href {https://doi.org/10.1007/JHEP01(2016)104}
  {\path{doi:10.1007/JHEP01(2016)104}}.

\bibitem{NEXT:2020jmz}
M.~Kekic, et~al., {Demonstration of background rejection using deep
  convolutional neural networks in the NEXT experiment}, JHEP 01 (2021) 189.
\newblock \href {http://arxiv.org/abs/2009.10783} {\path{arXiv:2009.10783}},
  \href {https://doi.org/10.1007/JHEP01(2021)189}
  {\path{doi:10.1007/JHEP01(2021)189}}.

\bibitem{Nygren:2009zz}
D.~Nygren, {High-pressure xenon gas electroluminescent TPC for $0-\nu ~ \beta
  \beta$-decay search}, Nucl.Instrum.Meth. A603 (2009) 337--348.
\newblock \href {https://doi.org/10.1016/j.nima.2009.01.222}
  {\path{doi:10.1016/j.nima.2009.01.222}}.

\bibitem{Avignone2008}
F.~T. Avignone~III, S.~R. Elliott, J.~Engel, {Double beta decay, Majorana
  neutrinos, and neutrino mass}, Rev. Mod. Phys. 80 (2008) 481.
\newblock \href {http://arxiv.org/abs/0708.1033} {\path{arXiv:0708.1033}},
  \href {https://doi.org/10.1103/RevModPhys.80.481}
  {\path{doi:10.1103/RevModPhys.80.481}}.

\bibitem{Dolinski2019}
M.~J. Dolinski, A.~W. Poon, W.~Rodejohann, {Neutrinoless Double-Beta Decay:
  Status and Prospects}, Ann. Rev. Nucl. Part. Sci. 69 (2019) 219--251.
\newblock \href {http://arxiv.org/abs/1902.04097} {\path{arXiv:1902.04097}},
  \href {https://doi.org/10.1146/annurev-nucl-101918-023407}
  {\path{doi:10.1146/annurev-nucl-101918-023407}}.

\bibitem{Gomez-Cadenas:2023vca}
J.~J. G\'omez-Cadenas, J.~Mart\'\i{}n-Albo, J.~Men\'endez, M.~Mezzetto,
  F.~Monrabal, M.~Sorel, {The search for neutrinoless double-beta decay}, Riv.
  Nuovo Cim. 46 (2023) 619--692.
\newblock \href {https://doi.org/10.1007/s40766-023-00049-2}
  {\path{doi:10.1007/s40766-023-00049-2}}.

\bibitem{Bilenky_1987}
S.~M. Bilenky, S.~T. Petcov, {Massive neutrinos and neutrino oscillations},
  Rev. Mod. Phys. 59 (1987) 671--754.
\newblock \href {https://doi.org/10.1103/RevModPhys.59.671}
  {\path{doi:10.1103/RevModPhys.59.671}}.

\bibitem{Aprile:2008bga}
E.~Aprile, A.~E. Bolotnikov, A.~L. Bolozdynya, T.~Doke, {Noble Gas Detectors},
  Wiley, 2008.
\newblock \href {https://doi.org/10.1002/9783527610020}
  {\path{doi:10.1002/9783527610020}}.

\bibitem{freitas_2010}
E.~Freitas, et~al., Secondary scintillation yield in high-pressure xenon gas
  for neutrinoless double beta decay (0$\nu\beta\beta$) search, Phys. Lett. B
  684 (2010) 205--210.
\newblock \href {https://doi.org/10.1016/j.physletb.2010.01.013}
  {\path{doi:10.1016/j.physletb.2010.01.013}}.

\bibitem{leardini_2022}
S.~Leardini, et~al., Time and band-resolved scintillation in time projection
  chambers based on gaseous xenon, Eur. Phys. J. C 82 (2022) 425.
\newblock \href {http://arxiv.org/abs/2112.04750} {\path{arXiv:2112.04750}},
  \href {https://doi.org/10.1140/epjc/s10052-022-10385-y}
  {\path{doi:10.1140/epjc/s10052-022-10385-y}}.

\bibitem{Renner:2019pfe}
J.~Renner, et~al., {Energy calibration of the NEXT-White detector with 1\%
  resolution near Q$_{\beta \beta}$ of $^{136}$Xe}, JHEP 10 (2019) 230.
\newblock \href {http://arxiv.org/abs/1905.13110} {\path{arXiv:1905.13110}},
  \href {https://doi.org/10.1007/JHEP10(2019)230}
  {\path{doi:10.1007/JHEP10(2019)230}}.

\bibitem{PhysRev.72.26}
U.~Fano, Ionization yield of radiations. ii. the fluctuations of the number of
  ions, Phys. Rev. 72 (1947) 26--29.
\newblock \href {https://doi.org/10.1103/PhysRev.72.26}
  {\path{doi:10.1103/PhysRev.72.26}}.

\bibitem{NEXT:2025yqw}
C.~Adams, et~al., {The NEXT-100 Detector}, Eur. Phys. J. C 86 (2026) 114.
\newblock \href {http://arxiv.org/abs/2505.17848} {\path{arXiv:2505.17848}},
  \href {https://doi.org/10.1140/epjc/s10052-025-14951-y}
  {\path{doi:10.1140/epjc/s10052-025-14951-y}}.

\bibitem{NEXT_2025_energy}
M.~P. Maneiro, et~al., Demonstration of sub-percent energy resolution in the
  next-100 detector (2025).
\newblock \href {http://arxiv.org/abs/2511.02467} {\path{arXiv:2511.02467}}.

\bibitem{Alvarez:2013gxa}
V.~Alvarez, et~al., {Operation and first results of the NEXT-DEMO prototype
  using a silicon photomultiplier tracking array}, JINST 8 (2013) P09011.
\newblock \href {http://arxiv.org/abs/1306.0471} {\path{arXiv:1306.0471}},
  \href {https://doi.org/10.1088/1748-0221/8/09/P09011}
  {\path{doi:10.1088/1748-0221/8/09/P09011}}.

\bibitem{Lorca:2014sra}
D.~Lorca, et~al., {Characterisation of NEXT-DEMO using xenon K$_{\alpha}$
  X-rays}, JINST 9 (2014) P10007.
\newblock \href {http://arxiv.org/abs/1407.3966} {\path{arXiv:1407.3966}},
  \href {https://doi.org/10.1088/1748-0221/9/10/P10007}
  {\path{doi:10.1088/1748-0221/9/10/P10007}}.

\bibitem{Alvarez:2012xda}
V.~\'Alvarez, et~al., {Initial results of NEXT-DEMO, a large-scale prototype of
  the NEXT-100 experiment}, JINST 8 (2013) P04002.
\newblock \href {http://arxiv.org/abs/1211.4838} {\path{arXiv:1211.4838}},
  \href {https://doi.org/10.1088/1748-0221/8/04/P04002}
  {\path{doi:10.1088/1748-0221/8/04/P04002}}.

\bibitem{NEXT_PRX_2022}
C.~A.~O. Henriques, et~al., Neutral bremsstrahlung emission in xenon unveiled,
  Phys. Rev. X 12 (2022) 021005.
\newblock \href {http://arxiv.org/abs/2202.02614} {\path{arXiv:2202.02614}},
  \href {https://doi.org/10.1103/PhysRevX.12.021005}
  {\path{doi:10.1103/PhysRevX.12.021005}}.

\bibitem{NEXT_TPB_2012}
V.~Álvarez, et~al., {SiPMs coated with TPB: coating protocol and
  characterization for NEXT}, JINST 7 (2012) P02010.
\newblock \href {https://doi.org/10.1088/1748-0221/7/02/P02010}
  {\path{doi:10.1088/1748-0221/7/02/P02010}}.

\bibitem{IC}
{NEXT Collaboration}, {IC: Invisible Cities},
  \url{https://github.com/next-exp/IC}.

\bibitem{Bolotnikov_1993}
A.~Bolotnikov, B.~Ramsey, Studies of light and charge produced by
  alpha-particles in high-pressure xenon, Nucl. Instrum. Meth. A 428~(2) (1999)
  391--402.
\newblock \href {https://doi.org/https://doi.org/10.1016/S0168-9002(99)00173-4}
  {\path{doi:https://doi.org/10.1016/S0168-9002(99)00173-4}}.

\bibitem{Platzman1961}
R.~Platzman, Total ionization in gases by high-energy particles: An appraisal
  of our understanding, Intern. J. Appl. Rad. Iso. 10 (1961) 116--127.
\newblock \href {https://doi.org/https://doi.org/10.1016/0020-708X(61)90108-9}
  {\path{doi:https://doi.org/10.1016/0020-708X(61)90108-9}}.

\bibitem{Henriques_2024}
C.~Henriques, J.~Teixeira, P.~Silva, R.~Mano, J.~dos Santos, C.~Monteiro,
  Understanding the xenon primary scintillation yield for cutting-edge rare
  event experiments, JCAP 2024 (2024) 041.
\newblock \href {http://arxiv.org/abs/2309.14202} {\path{arXiv:2309.14202}},
  \href {https://doi.org/10.1088/1475-7516/2024/06/041}
  {\path{doi:10.1088/1475-7516/2024/06/041}}.

\bibitem{NEXT:2023blw}
K.~Mistry, et~al., {Design, characterization and installation of the NEXT-100
  cathode and electroluminescence regions}, JINST 19 (2024) P02007.
\newblock \href {http://arxiv.org/abs/2311.03528} {\path{arXiv:2311.03528}},
  \href {https://doi.org/10.1088/1748-0221/19/02/P02007}
  {\path{doi:10.1088/1748-0221/19/02/P02007}}.

\bibitem{Sorensen_2025}
P.~Sorensen, R.~Gibbons, Quartz fluorescence backgrounds in xenon particle
  detectors, Phys. Rev. D 112 (2025) 052004.
\newblock \href {http://arxiv.org/abs/2505.08067} {\path{arXiv:2505.08067}},
  \href {https://doi.org/10.1103/lsmf-pm5g} {\path{doi:10.1103/lsmf-pm5g}}.

\end{thebibliography}

\end{document}